\documentclass[twocolumn,epjc3]{svjour3}
\usepackage{amsmath,amssymb}
\usepackage{graphics}
\usepackage[dvipsnames]{xcolor}
\usepackage{cuted}
\usepackage{microtype}
\usepackage{txfonts}
\usepackage{bbold}
\usepackage{anyfontsize}
\usepackage[pdftex]{graphicx}
\usepackage{nicematrix,tikz}
\usepackage{pgfplots}
\pgfplotsset{width=10cm,compat=1.9}
\usepackage{xargs}       
\usepackage{enumerate}
\usepackage{appendix}
\usepackage{float}
\usepackage{stmaryrd}
\usepackage[parfill]{parskip}
\usepackage[hyperindex=true,colorlinks=true]{hyperref}
\usepackage{titletoc}
\usepackage{makecell}
\usepackage{lscape}
\usepackage{blkarray}
\usepackage{multirow}
\usetikzlibrary{plotmarks}
\usepackage{braket}
\usepackage{cancel}
\usepackage[export]{adjustbox}
\usepackage[sort&compress,numbers]{natbib}

\allowdisplaybreaks

\hypersetup{
    citecolor=blue,
    linkcolor=blue,
    urlcolor=blue
} 

\begin{document}
\title{\textit{Ab initio} description of monopole resonances in light- and medium-mass nuclei}
\subtitle{IV. Angular momentum projection and rotation-vibration coupling}
\author{A. Porro\thanksref{ad:tud,ad:emmi,ad:saclay} 
\and T. Duguet\thanksref{ad:saclay,ad:kul}
\and J.-P. Ebran\thanksref{ad:dam,ad:dam_u}
\and M. Frosini\thanksref{ad:cadarache}
\and R. Roth\thanksref{ad:tud,ad:darm2}
\and V. Som\`a\thanksref{ad:saclay}}

\institute{
\label{ad:tud}
Technische Universit\"at Darmstadt, Department of Physics, 64289 Darmstadt, Germany
\and
\label{ad:emmi}
ExtreMe Matter Institute EMMI, GSI Helmholtzzentrum f\"ur Schwerionenforschung GmbH, 64291 Darmstadt, Germany
\and
\label{ad:saclay}
IRFU, CEA, Universit\'e Paris-Saclay, 91191 Gif-sur-Yvette, France 
\and
\label{ad:kul}
KU Leuven, Department of Physics and Astronomy, Instituut voor Kern- en Stralingsfysica, 3001 Leuven, Belgium
\and
\label{ad:dam}
CEA, DAM, DIF, 91297 Arpajon, France
\and
\label{ad:dam_u}
Universit\'e Paris-Saclay, CEA, Laboratoire Mati\`ere en Conditions Extr\^emes, 91680 Bruy\`eres-le-Ch\^atel, France
\and
\label{ad:cadarache}
CEA, DES, IRESNE, DER, SPRC, 13108 Saint-Paul-l\`es-Durance, France
\and
\label{ad:darm2}
Helmholtz Forschungsakademie Hessen f\"ur FAIR, GSI Helmholtzzentrum, 64289 Darmstadt, Germany
}
\date{Received: \today{} / Revised version: date}

\maketitle
%
%
\begin{abstract}
Giant Resonances are, with nuclear rotations, the most evident expression of collectivity in finite nuclei. These two categories of excitations, however, are traditionally described within different formal schemes, such that vibrational and rotational degrees of freedom are separately treated and coupling effects between those are often neglected. The present work puts forward an approach aiming at a consitent treatment of vibrations and rotations. Specifically, this paper is the last in a series of four dedicated to the investigation of the giant monopole resonance in doubly open-shell nuclei via the ab initio Projected Generator Coordinate Method (PGCM). The present focus is on the treatment and impact of angular momentum restoration within such calculations. The PGCM being based on the use of deformed mean-field states, the angular-momentum restoration is performed when solving the secular equation to extract vibrational excitations. In this context, it is shown that performing the angular momentum restoration only after solving the secular equation contaminates the monopole response with an unphysical coupling to the rotational motion, as was also shown recently for (quasi-particle) random phase approximation calculations based on a deformed reference state. Eventually, the present work based on the PGCM confirms that an {\it a priori} angular momentum restoration  is necessary to handle consistently both collective motions at the same time. This further pleads in favor of implementing the full-fledged projected (quasi-particle) random phase approximation in the future.
\end{abstract}

\section{Introduction}

Giant Resonances (GRs)~\cite{Berman75,harakeh2001giant,Garg2018a,Colo:2022vmu} are collective nuclear excitations that can best be pictured in terms of oscillations of the nuclear surface in an effective liquid-drop model. GRs are, in this respect, one of the clearest manifestations of collective motion in finite nuclei. The other most evident collective behaviour in nuclei is provided by rotational excitations~\cite{Elliott58}. These features have traditionally been described through empirical models adopting various resolutions, e.g. the macroscopic Bohr-Mottelson collective models~\cite{Bohr98a} and their microscopic counterparts~\cite{Rowe_1985,Libert16}, the Interacting Boson Model~\cite{ARIMA76,ARIMA78,Iachello_Arima_1987}, the Fermion Dynamical Symmetry Model~\cite{FDSM}, etc. In a microscopic framework, a unified and rigorous treatment of both rotations and vibrations is not amenable to simple solutions of the Schrödinger equation.
For this reason, the two are often addressed separately.
This is justified by the fact that nuclear rotations and vibrations typically pertain to different energy regimes.
Indeed, rotations are low-energy excitations, with ground-state rotational bands typically spanning few MeV's above the nuclear ground state. Instead, GRs typically appear above 10 MeV excitation energy with large differences depending on the system and on the multipolarity of interest. 

When addressing doubly open-shell nuclei, both realms are typically approached starting from mean-field reference states breaking rotational symmetry, e.g. deformed Hartree-Fock-Bogoliubov (HFB) vacua. Rotations are then extracted by performing an angular momentum projection (AMP) acting either on one deformed HFB vacuum~\cite{SHEIKH00} or on a linear superposition of them, thus defining the projected generator coordinate method (PGCM)~\cite{bender03b,Robledo_2019}. 
Lately, rotational spectra have also been generated starting from deformed coupled cluster (CC) calculations with explicit AMP~\cite{Hagen:2022tqp,Sun:2024iht,Hu:2024pee}. 
As for GRs, the (quasi-particle) random phase approximation ((Q)RPA) is the usual method of choice, where nuclear vibrations are treated as harmonic fluctuations around the deformed HFB minimum~\cite{Peru08,Yoshida2009,Avogadro11a, Beaujeault23a}, without any AMP. 

In a recent work~\cite{Porro23p}, an AMP in RPA calculations was attempted {\it a posteriori}, i.e. after solving the RPA secular equation based on the deformed reference state. The corresponding monopole response was shown to be contaminated with an unphysical coupling to the rotational motion and an empirical method was designed to subtract it. On a conceptually similar ground, when $U(1)$ symmetry is broken, QRPA pair transfer probabilities have been demonstrated to overestimate the exact results within the exactly solvable Richardson model~\cite{richardson1964exact}. The most significant discrepancies occur near the transition from the normal to the superfluid phase~\cite{Gambacurta:2012tf}. On the other hand, the present series of four papers~\cite{Porro24a,Porro24b,Porro24c} addressing the giant monopole resonance (GMR) from an \textit{ab initio} standpoint has demonstrated the suitability of the symmetry-conserving PGCM to address GRs. In this context, the goal of the present paper, the fourth of the series, is to investigate whether performing the AMP {\it a posteriori} on top of the GCM solutions induces the same shortcomings as those observed within the RPA framework and thus confirm that the {\it a priori} AMP is mandatory to properly handling the coupling between rotational and vibrational motions. 

The present paper, denoted as Paper IV, is organized as follows\footnote{The first three papers of the series are denoted as Paper I~\cite{Porro24a}, Paper II~\cite{Porro24b} and Paper III~\cite{Porro24c}, respectively.}. First, the different strategies to perform a symmetry restoration within the (P)GCM are formally introduced in Sec.~\ref{sec:theo} before discussing the impact on the strength function in Sec.~\ref{sec:GCM_proj_levels}. In Sec.~\ref{sec:spurious_rot}, the empirical method used to remove the spurious coupling due to the {\it a posteriori} AMP is detailed. Numerical results are eventually presented in Sec.~\ref{chap:proj_GCM_res} whereas conclusions are drawn in Sec.~\ref{sec:pav_gcm_summary} .

\section{Formalism}
\label{sec:theo}

The (P)GCM formalism was introduced in detail in Paper I~\cite{Porro24a}. The essential elements  needed to discuss the effects of the symmetry breaking and restoration in strength functions are only briefly recalled below. 

\subsection{The generator coordinate method}
\label{sec:GCM_theo}

The original GCM formulation is introduced first. The GCM wave-function ansatz~\cite{Hill53a,Griffin57a} is a general superposition of so-called generating functions reading as
\begin{equation}
\label{eq:GCM_ansatz}
    \ket{\Psi_\nu}\equiv\sum_q f_\nu(q)\ket{\Phi(q)}\,,
\end{equation}
where $q$ denotes a set of variables referred to as \textit{the generator coordinates}. The index $\nu$ refers to a principal quantum number while $f_\nu(q)$ is a weight function to be determined\footnote{The mixing coefficients $\{f_\nu(q),\,q\in[q_0,q_1]\}$ are defined such that $\ket{\Psi_\nu}$ is normalized.}. The ensemble $\{ | \Phi(q),\,q\in[q_0,q_1] \rangle\}$  denotes a set of non-orthogonal Bogoliubov states typically obtained as solutions of constrained HFB calculations requiring that the solution satisfies
\begin{align}
\langle \Phi(q) | Q | \Phi(q) \rangle &= q \, , \label{constraint}
\end{align}
where $Q$ denotes a set of operators defining the collective coordinates. 

The unknown coefficients $f_\nu(q)$ are determined variationally based on Ritz' variational principle, namely by minimising the energy associated with $\ket{\Psi_\nu}$
\begin{equation}
	\delta\frac{\braket{\Psi_\nu|H|\Psi_\nu}}{\braket{\Psi_\nu|\Psi_\nu}}=0.
\end{equation}
The variation with respect to the weights $f_\nu^{*}(q)$ eventually leads to a generalised eigenvalue problem known as the \textit{Hill-Wheeler-Griffin} (HWG) secular equation~\cite{Griffin57a} reading as
\begin{equation}
\label{eq:HWG}
	\sum_q\Big[\mathcal{H}(p,q)-E_\nu\mathcal{N}(p,q)\Big]f_\nu(q)=0\,,
\end{equation}
where the so-called Hamiltonian and norm kernels are respectively defined as
\begin{subequations}
    \label{eq:kernels}
	\begin{align}
		\mathcal{H}(p,q)&\equiv\braket{\Phi(p)|H|\Phi(q)}\,,\\
		\mathcal{N}(p,q)&\equiv\braket{\Phi(p)|\Phi(q)}\,.
	\end{align}
\end{subequations}

\subsection{The projected generator coordinate method}
\label{sec:PGCM_theo}

Constrained HFB solutions typically break symmetries of the Hamiltonian, so that restoring those symmetries is mandatory to obtain approximations to exact eigentates carrying good symmetry quantum numbers. The symmetry restoration is achieved within the PGCM by adding a projection operator onto good symmetry quantum numbers to the symmetry-breaking GCM state. While presently focusing on rotational symmetry associated with angular-momentum conservation, the approach is general and consists of modifying Eq. \eqref{eq:GCM_ansatz} according to\footnote{The mixing coefficients $\{f^\sigma_\nu(q),\,q\in[q_0,q_1]\}$ are defined such that $\ket{\Psi^\sigma_\nu}$ is normalized.}
\begin{equation}\label{eq:PGCM_ansatz1}
	\ket{\Psi_\nu^{\sigma}}=\sum_q\,f_\nu^\sigma(q)P^\sigma\ket{\Phi(q)}\,,
\end{equation}
where $P^\sigma$ is the projection operator associated with the (a) symmetry (sub)group $\mathcal{G}$ of the Hamiltonian. The projection operator selects the components of each $\ket{\Phi(q)}$ carrying the good symmetry quantum numbers $\sigma\equiv(JM\Pi NZ)$, i.e. the PGCM ansatz in Eq.~\eqref{eq:PGCM_ansatz1} has good total angular momentum $J$ and angular momentum projection $M$, parity $\Pi=\pm1$ as well as neutron $N$ and proton $Z$ numbers.
The projector $P^\sigma$ can be generically written as 
\begin{equation}
	P^\sigma=\int\text{d}\varphi\,g^\sigma(\varphi)R(\varphi)\,,
\end{equation} 
where $g^\sigma(\varphi)$ represents the irreducible representations of $\mathcal{G}$ whereas $R(\varphi)$ denotes the unitary symmetry transformation operator changing the orientation of a state by an angle $\varphi$. The PGCM ansatz can thus be expanded as
\begin{align}
\label{eq:PGCM_ansatz}
\ket{\Psi_\nu^{\sigma}}&=\sum_q\,f_\nu^\sigma(q)\int\text{d}\varphi\,g^\sigma(\varphi)R(\varphi)\ket{\Phi(q)}\nonumber\\
	&\equiv\sum_q\,f_\nu^\sigma(q)\int\text{d}\varphi\,g^\sigma(\varphi)\ket{\Phi(q,\varphi)}\,,
\end{align}
where the $\varphi$-rotated Bogoliubov state $\ket{\Phi(q,\varphi)}\equiv~R(\varphi)\ket{\Phi(q)}$ has been introduced. Applying the variational procedure based on the PGCM ansatz (Eq.~\eqref{eq:PGCM_ansatz}) leads now to a set of $\sigma$-dependent HWG equations
\begin{equation}\label{eq:P_HWG}
	\sum_q\,\Big[\mathcal{H}^\sigma(p,q)-E^\sigma_\nu\mathcal{N}^\sigma(p,q)\Big]f_\nu^\sigma(q)=0\,,
\end{equation}
where the so-called symmetry-restored Hamiltonian and norm kernels are defined as
\begin{subequations}\label{eq:P_kernels}
	\begin{align}
		\mathcal{H}^\sigma(p,q)&\equiv\braket{\Phi(p)|HP^\sigma|\Phi(q)}\nonumber\\
		&=\int\text{d}\varphi\,g^\sigma(\varphi)\braket{\Phi(p)|H|\Phi(q,\varphi)}\,,\\
		\mathcal{N}^\sigma(p,q)&\equiv\braket{\Phi(p)|P^\sigma|\Phi(q)}\nonumber\\
		&=\int\text{d}\varphi\,g^\sigma(\varphi)\braket{\Phi(p)|\Phi(q,\varphi)}\,.
	\end{align}
\end{subequations}

\subsection{PAV-GCM}
\label{sec:PAV_GCM_theo}

In the PGCM described above the secular equation (Eq.~\eqref{eq:P_HWG}) is solved in presence of the symmetry projection, i.e. the variational minimization is restricted to each irreducible representation of the symmetry group. For this reason, this scheme is presently denoted as the \textit{variation after projection} GCM (VAP-GCM).

In between the GCM and the VAP-GCM, a scheme can be considered in which the symmetry projection is performed only {\it after} the GCM solution based on a symmetry breaking ansatz has been obtained. Such an intermediate approach is naturally denoted as the \textit{projection after variation} GCM (PAV-GCM) scheme. 

Projecting the GCM states solution of Eqs.~\eqref{eq:GCM_ansatz}-\eqref{eq:HWG}, one works with the set of projected states
\begin{align}
	\label{eq:state_PAV_GCM}
	\ket{\Tilde{\Psi}^\sigma_\nu}&\equiv N_\nu^\sigma P^\sigma\ket{\Psi_\nu} =N_\nu^\sigma\sum_q f_\nu(q)P^\sigma\ket{\Phi(q)}\,,
\end{align}
where $N_\nu^\sigma$ is a normalising factor provided by the condition
\begin{align}
	1&=\braket{\Tilde{\Psi}_\nu^\sigma|\Tilde{\Psi}_\nu^\sigma} \,,
\end{align}
such that
\begin{equation}
	(N_\nu^{\sigma})^{-2}=\sum_{pq}f_\nu^*(p)f_\nu(q)\mathcal{N}^\sigma(p,q)\,. \label{normfact}
\end{equation}

\subsection{Discussion} 

\setlength{\tabcolsep}{20pt}
\begin{table*}[ht]
	\centering
	\begin{NiceTabular}{c:c}
		\hline
        \hline
		\multicolumn{2}{c}{\textcolor{red}{\textbf{Symmetry breaking}}}\\
		\hline 
		\textbf{GCM} & \textbf{(Q)RPA} \\
		\makecell[c]{Large-amplitude superposition \\ of deformed HF(B) states} & \makecell[c]{Harmonic fluctuations \\ around a deformed HF(B) state}\\
		\textbf{Status:} available & \textbf{Status:} available \\
		\hline
		\textbf{PAV GCM} & \textbf{PAV (Q)RPA} \\
		\makecell[c]{Angular-momentum projection \\ of deformed GCM states} & \makecell[c]{Angular-momentum projection \\ of deformed (Q)RPA states} \\
		\textbf{Status:} developed in this work & \textbf{Status:} developed in Refs.~\cite{erler_phd,Porro_thesis,Porro23p}. \\
		\hline
		\textbf{PGCM} & \textbf{P(Q)RPA} \\
		\makecell[c]{Proper treatment of rotation-vibration \\ coupling within the GCM} & \makecell[c]{Proper treatment of rotation-vibration \\ coupling within the (Q)RPA} \\
		\textbf{Status:} available & \textbf{Status:} formalism available~\cite{FedRi85,Tsuchimochi15a} \\
		\hline
		\multicolumn{2}{c}{\textcolor{Green}{\textbf{Symmetry conserving}}}\\
		\hline
        \hline
	\end{NiceTabular}
	\caption{Schematic representation of different projection levels based on symmetry-breaking GCM and (Q)RPA.}
	\label{tab:projection_scheme}
\end{table*}

The PAV strategy assumes that rotational and vibrational degrees of freedom are strictly decoupled, i.e. intrinsically deformed solutions are first obtained assuming that rotational degrees of freedom are frozen before adding the rotational motion to each vibrational state thus obtained. 

From a physical standpoint, such a decoupling presuppose that vibrations and rotations relate to very different time scales, such that they can be addressed separately in a Born-Oppenheimer-like approximation. Specifically, rotations are assumed to be infinitely slower than nuclear vibrations, which is a direct consequence of the rotation being ideally associated with a zero-energy (Goldstone) mode. 

However, nuclear rotations happen in fact at finite frequencies, such that they cannot be decoupled \textit{a priori} from vibrational modes~\cite{Bohr98a}. Thus, the variational/diagonalisation process at play to determine physical states should be performed in a Hilbert subspace simultaneously accounting for vibrational \textit{and} rotational degrees of freedom. This is achieved in the VAP scheme that is the method of choice to consistently treat the coupling effects between nuclear vibrations and rotations. A schematic summary of the different levels of symmetry breaking and restoration in the GCM and the (Q)RPA is displayed in Table~\ref{tab:projection_scheme}.

\section{Transition strength}
\label{sec:GCM_proj_levels}

Generically, the ground-state strength function associated with an arbitrary excitation operator $O$ reads as
\begin{equation}
    S_{F}(\omega)\equiv \sum_{\nu \sigma'} |\braket{\Theta^{\sigma_0}_0|O|\Theta^{\sigma'}_\nu}|^2 \, \delta(E^{\sigma'}_\nu-E^{\sigma_0}_0-\omega) \, , \label{strengthfunction}
\end{equation}
where $| \Theta^{\sigma}_\nu \rangle$ ($E^{\sigma}_\nu$) denotes an eigenstate (eigenenergy) of the nuclear Hamiltonian. 

The present work focuses on the monopole response and on the $K=0$ component of the quadrupole response, respectively associated with the excitation operators
\begin{align}
O &\equiv r^2 \equiv \sum_{i=1}^\text{A} r_i ^2  \label{msrop} \, . 
\end{align}
and
\begin{align}
O &\equiv Q_{20} \equiv \sum_{i=1}^A r_i^2 \, Y_{20}(\vartheta_i, \phi_i) \, , \label{quadop}
\end{align}
where $(\vartheta_i, \phi_i)$ denote spherical angular coordinates. Below, the ingredients entering Eq.~\eqref{strengthfunction} in GCM, PAV-GCM and VAP-GCM calculations are specified.

\subsection{GCM}
\label{sec:trans_GCM}

The unprojected transition amplitude between the GCM ground state and a GCM excited state reads as
\begin{align}
	\braket{\Psi_0|O|\Psi_\nu}&=\sum_{pq}f_0^*(p)f_\nu(q)\braket{\Phi(p)|O|\Phi(q)}\nonumber\\
	&\equiv\sum_{pq}f_0^*(p)O(p,q)f_\nu(q)\,. \label{eq:trans_GCM0}
\end{align}
Using more compact notations, the transition matrix element between any two GCM states can be written as
\begin{align}
	\label{eq:trans_GCM}
	\braket{\Psi_\mu|O|\Psi_\nu}&\equiv\sum_{pq}f^*_{\mu p}O_{pq}f_{q\nu} \equiv(\textbf{f}^\dagger\cdot\textbf{O}\cdot\textbf{f})_{\mu\nu}\,,
\end{align}
where the indices of the matrix $\textbf{O}$ associated with the operator kernel $O(p,q)$ run over the generator coordinates whereas the linear coefficient matrix $\textbf{f}$ indices run on the generator coordinates for the lines and on the GCM states for the columns. Naturally, the energies entering Eq.~\eqref{strengthfunction} are the GCM energies delivered by Eq.~\eqref{eq:HWG}.

\subsection{VAP-GCM}
\label{sec:trans_PGCM}

In PGCM calculations, the transition amplitude reads as
\begin{align}
	\braket{\Psi_0^{\sigma_0}|O|\Psi_\nu^{\sigma}}&=\sum_{pq}f_0^{\sigma_0*}(p)f_\nu^{\sigma}(q)\braket{\Phi(p)|P^{\sigma_0\dagger}OP^{\sigma}|\Phi(q)} \nonumber \\
 &\equiv \sum_{pq}f_0^{\sigma_0*}(p) O^{\sigma_0\sigma}(p,q)f_\nu^{\sigma}(q) \nonumber \\ 
&\equiv\sum_{pq}f_{0 p}^{\sigma_0*}O_{pq}^{\sigma_0\sigma}f_{q\nu}^{\sigma}\nonumber\\
&\equiv(\textbf{f}^{\sigma_0\dagger}\cdot\textbf{O}^{\sigma_0\sigma}\cdot\textbf{f}^{\sigma})_{0\nu}\,,
 \label{eq:trans_VAP_GCM0}
\end{align}
where the matrix $\textbf{O}^{\sigma\sigma'}$ associated with the projected kernel $O^{\sigma\sigma'}(p,q)$ carries the symmetry quantum numbers of both the bra and ket. Indeed, and contrary to the Hamiltonian, the operator $O$ is not necessarily a scalar under symmetry transformations.  Naturally, the energies entering Eq.~\eqref{strengthfunction} are the PGCM energies delivered by Eq.~\eqref{eq:P_HWG}.

\subsection{PAV-GCM}

Given the PAV-GCM states introduced in Eq.~\eqref{eq:state_PAV_GCM}, the corresponding transition amplitude reads as
\begin{align}
\braket{\Tilde{\Psi}^{\sigma_0}_0|O|\Tilde{\Psi}^{\sigma}_\nu} & =N^{\sigma_0}_0N^{\sigma}_\nu\sum_{pq}f_0(p)f_\nu(q)\braket{\Phi(p)|P^{\sigma_0\dagger}OP^{\sigma}|\Phi(q)} \nonumber \\
&=N^{\sigma_0}_0 N^{\sigma}_\nu 
 (\textbf{f}^{\dagger}\cdot\textbf{O}^{\sigma_0\sigma}\cdot\textbf{f})_{0\nu}\,. \label{eq:trans_PAV_GCM}
\end{align}
Up to a normalising factor, the PAV-GCM transition amplitude combines the projected kernels introduced in Eq.~\eqref{eq:trans_VAP_GCM0} with the mixing coefficients of the two involved GCM states. 

In this scheme, the energies entering Eq.~\eqref{strengthfunction} are the PAV-GCM energies delivered by 
\begin{align}
	\Tilde{E}^{\sigma}_\nu &\equiv \frac{\braket{\Tilde{\Psi}^{\sigma}_\nu|H|\Tilde{\Psi}^{\sigma}_\nu}}{\braket{\Tilde{\Psi}^{\sigma}_\nu|\Tilde{\Psi}^{\sigma}_\nu}} = |N^{\sigma}_\nu|^2
 (\textbf{f}^{\dagger}\cdot\textbf{H}^{\sigma\sigma}\cdot\textbf{f})_{\nu\nu} \, ,
\end{align}
where in fact a single projector is sufficient to compute the kernel given that the Hamiltonian is a scalar under rotation.

\section{Spurious coupling to rotational motion}
\label{sec:spurious_rot}

In this section the concept of spurious coupling to the rotational motion in theories breaking angular-momentum conservation is introduced. For the reasons mentioned in Sec.~\ref{sec:PGCM_theo}, vibrational GCM excitations obtained without AMP may be non-orthogonal to rotational states. This feature is considered spurious given that the neglect of the rotational degrees of freedom in the HWG equation precisely assumes that the intrinsic GCM states are fully decoupled from them. Based on this consideration, a method is now designed to subtract {\it a posteriori} the spurious coupling between excited GCM states and a pure rotational motion of the corresponding ground state. 

\subsection{Subtracted GCM}

Given the overlap between an arbitrary GCM excited state $\ket{\Psi_\nu}$ and the PAV-GCM ground state 
\begin{align}
	\label{eq:a_rot}
	a_\nu&\equiv\braket{\Tilde{\Psi}^{\sigma_0}_0|\Psi_\nu} \, ,
\end{align}
the spurious coupling can be subtracted by redefining the excited state\footnote{The procedure can be applied to any many-body method accessing symmetry-breaking ground and excited states.} as~\cite{Faessler88a}
\begin{equation}
	\label{eq:prescr_sub_rem}
    \ket{\breve{\Psi}_\nu}\equiv N_{\breve{\nu}}\big[\ket{\Psi_\nu}-a_\nu\ket{\Tilde{\Psi}^{\sigma_0}_0}\big]\,.
\end{equation}
It is immediate to check that the orthogonalisation condition 
\begin{equation}
	\label{eq:ortho_cond}
    \braket{\Tilde{\Psi}^{\sigma_0}_0|\breve{\Psi}_\nu}=0 \, 
\end{equation}
is satisfied and that the normalisation constant $N_{\breve{\nu}}$ is given by
\begin{equation}
    (N_{\breve\nu})^{-2}=1-|a_\nu|^2\,.
\end{equation}
Replacing $\ket{\Psi_\nu}$ by $\ket{\breve{\Psi}_\nu}$ in Eq.~\eqref{eq:trans_GCM0}, the subtracted GCM (sub-GCM) transition strength can be computed.

\subsection{Subtracted PAV-GCM}

The subtraction method is extended to the PAV-GCM via the introduction of 
\begin{equation}
	\label{eq:state_sub_PAV_GCM}
	\ket{\breve{\Tilde{\Psi}}_\nu^{\sigma}}\equiv N_{\breve{\nu}}^{\sigma}P^{\sigma}\ket{\breve{\Psi}_\nu}\,,
\end{equation}
where the coefficients $a_\nu$ are now defined to fulfill the orthogonalisation condition
\begin{equation}
	\braket{\Tilde{\Psi}^{\sigma_0}_0|\breve{\Tilde{\Psi}}_\nu^{\sigma}}=0\,.
\end{equation}
In fact, $\ket{\breve{\Tilde{\Psi}}_\nu^{\sigma}}$ differ from the PAV-GCM state $\ket{\Tilde{\Psi}_\nu^{\sigma}}$ only if it carries the same symmetry quantum numbers as the ground state, i.e.  $\sigma=\sigma_0$, which in the case of present interest corresponds to $J=0$ states. Correspondingly, the coefficient $a_\nu$ satisfies Eq.~\eqref{eq:a_rot} and the normalising factor $N_{\breve{\nu}}^{\sigma_0}$ reads
\begin{align}
	\label{eq:norm_sub_pav_rpa}
	(N_{\breve{\nu}}^{\sigma_0})^{-2}&=\braket{\Psi_\nu|P^{\sigma_0}|\Psi_\nu}-|a_\nu|^2=(N_\nu^{\sigma_0})^{-2}-|a_\nu|^2\,.
\end{align}
Replacing $\ket{\Tilde{\Psi}^{\sigma}_\nu}$ by $\ket{\breve{\Tilde{\Psi}}^{\sigma}_\nu}$ in Eq.~\eqref{eq:trans_PAV_GCM}, the subtracted PAV-GCM (sub-PAV-GCM) transition strength can be computed.

\section{Applications}
\label{chap:proj_GCM_res}

The above considerations are illustrated below in the case of $^{28}$Si that acts as a typical example. The general conclusions reached below have been checked to be valid in all the nuclei studied in Papers I, II and III. 

\subsection{Numerical setting} 

Calculations whose results are presented below are realised employing a spherical harmonic oscillator basis characterized by $\hbar\omega=12$ MeV and $\;e_{\!_{\;\text{max}}}=10$ and the chiral effective field theory ($\chi$EFT) Hamiltonian of Ref.~\cite{Hu20} built at next-to-next-to-next-to-leading order (N$^{3}$LO). Two-dimensional (P)GCM calculations are performed in axial symmetry using $(r,\beta_2)$ as generator coordinates. All details of the calculations can be found in Paper II~\cite{Porro24b}. Specifically, here particle-number projection is always included such that GCM calculations only omit the AMP, which is focus of the present work.

\subsection{Levels of calculation} 

The different levels of calculations discussed in the present work are 
\begin{itemize}
    \item \textbf{GCM}: no AMP.
    \item \textbf{PAV-GCM}: AMP performed \textit{a posteriori} on GCM states. 
    \item \textbf{VAP-GCM}: full PGCM with AMP \textit{a priori}.
    \item \textbf{sub-GCM}: GCM with \textit{a posteriori} subtraction of the rotational coupling.
    \item \textbf{sub-PAV-GCM}: AMP performed \textit{a posteriori} on  sub-GCM states.
\end{itemize}

\subsection{Results}

\begin{figure}
	\includegraphics[width=0.42\textwidth]{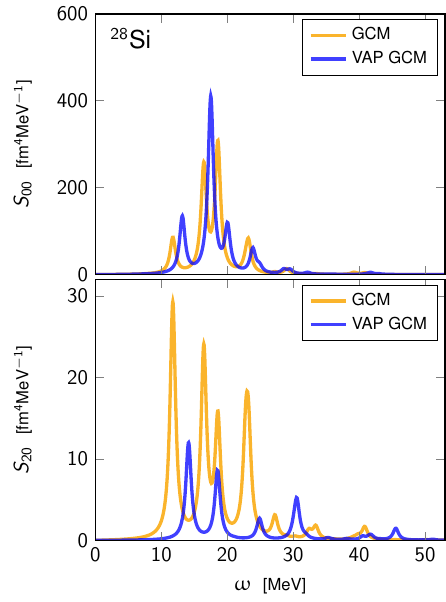}
	\caption{Monopole (top) and quadrupole (bottom) responses for GCM and VAP-GCM calculations in $^{28}$Si.}
	\label{fig:Si28_0}
\end{figure}

The monopole (quadrupole) GCM and VAP-GCM responses in $^{28}$Si are compared in the upper (lower) panel of Fig.~\ref{fig:Si28_0}. The GCM monopole response is fragmented among four peaks in the interval $[11,23]$\,MeV, the two dominant ones being located around $18$\,MeV. These four peaks are fully correlated with those appearing in the quadrupole response, even though their relative weights are different in the two cases. This correlation is the fingerprint of the coupling between both modes due to the intrinsic (oblate) deformation of $^{28}$Si. This topic was discussed at length in Paper II~\cite{Porro24b}. 

The inclusion of the AMP in VAP-GCM calculations impacts the responses in two ways. First, the excitation energy of the four dominant peaks are shifted up by $1.4$, $1.0$, $1.5$ and $0.7$\,MeV, respectively, in the $J=0$ (monopole) channel and by  $2.4$, $2.1$, $2.5$ and $1.7$\,MeV in the $J=2$ (quadrupole) channel, i.e. the $J=0$ and $J=2$ states originating from the same intrinsic state are no longer strictly degenerate. Second, the intensity of the peaks is modified. The quadrupole response is significantly suppressed, i.e. while the strength of the first two peaks is divided by about a factor of three, the third peak has entirely disappeared and the fourth peak has been severely shrunk\footnote{The third (fourth) peak visible in the VAP-GCM quadrupole response corresponds to the fourth (fifth) peak in its GCM counterpart.}. The intensity of the peaks in the monopole response remains overall unchanged except that the relative weight of the two main contributions near $18$\,MeV is strongly modified. Overall, the {\it a priori} inclusion of the AMP impacts the monopole and quadrupole responses non negligibly but it does so in a way that maintains a close connection to their {\it intrinsic} GCM counterparts. 

\begin{figure}
	\includegraphics[width=0.5\textwidth]{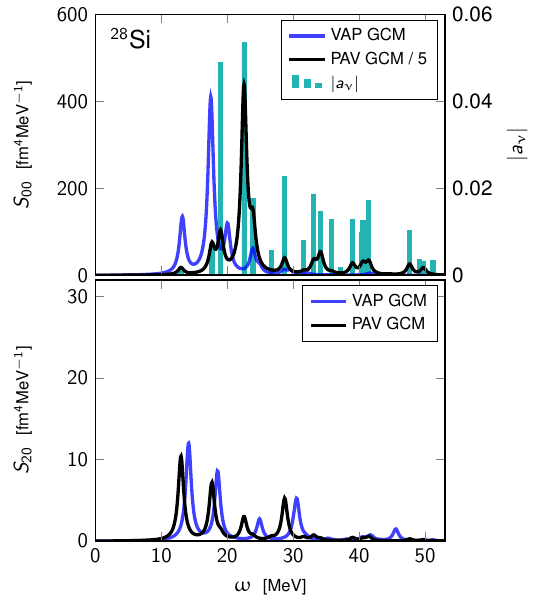}
	\caption{Monopole (top) and quadrupole (bottom) responses for VAP-GCM and PAV-GCM calculations in $^{28}$Si. For the monopole response, the overlap between excited GCM states and the PAV-GCM ground state is also displayed, the corresponding scale being shown on the right y-axis.}
	\label{fig:Si28_1}
\end{figure}

Next, VAP-GCM and PAV-GCM results are compared in Fig.~\ref{fig:Si28_1}. The quadrupole responses are very similar except for a shift up by about $1$\,MeV for the $J=2$ VAP-GCM excitation energies compared to the PAV-GCM ones. Contrarily, the monopole responses differ very notably. As a matter of fact, the {\it a posteriori} AMP impacts the monopole amplitudes much more significantly\footnote{Notice the rescaling of the PAV-GCM response by the factor $1/5$.} than in the VAP-GCM calculation\footnote{The $J=0$ energies are little affected, i.e. PAV-GCM and VAP-GCM energies typically differ by $0.5$\,MeV.}. This is particularly true beyond the first three peaks where the originally subleading fourth peak is very strongly enhanced, along with many significant peaks appearing at even higher energies where no GCM strength was visible in the first place. The anomalously large impact of the {\it a posteriori} AMP compared to the VAP-GCM results makes the validity of the PAV-GCM monopole response dubious. 

In order to analyse the content of these results, the overlap $a_\nu$ between excited GCM states and the PAV-GCM ground-state is also displayed in Fig.~\ref{fig:Si28_1}. The PAV-GCM monopole strength happens to be anomalously large\footnote{The overlap reaches $0.05$ in $^{28}$Si and is up to three times larger in other studied nuclei such as $^{46}$Ti or $^{24}$Mg. In the studied cases, the coupling is larger for low-energy (high-energy) states in prolate (oblate) nuclei.} for GCM excited states that are strongly coupled to the PAV-GCM ground state, even though no associated strength was originally present in the GCM response.

\begin{figure}
	\includegraphics[width=0.42\textwidth]{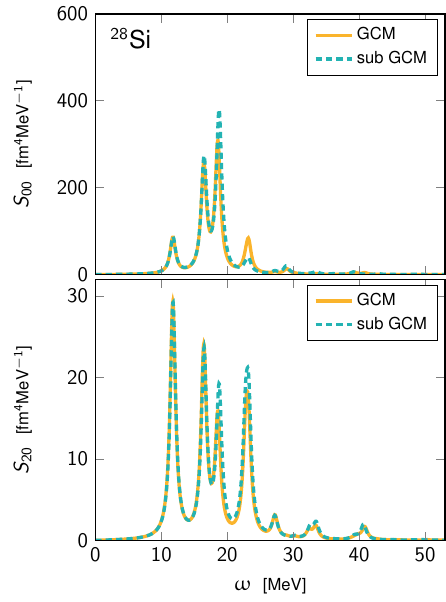}
	\caption{Monopole (top) and quadrupole (bottom) responses for GCM and sub-GCM calculations in $^{28}$Si.}
	\label{fig:Si28_2}
\end{figure}

\begin{figure}
	\includegraphics[width=0.42\textwidth]{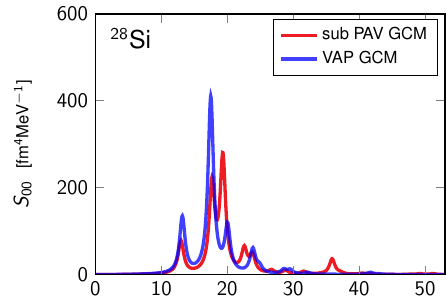}
	\caption{Monopole response for VAP-GCM and sub-PAV-GCM calculations in $^{28}$Si.}
	\label{fig:Si28_3}
\end{figure}

As discussed in Sec.~\ref{sec:spurious_rot}, the spurious coupling to the rotational motion can be subtracted {\it a posteriori} from GCM and PAV-GCM results. Figure~\ref{fig:Si28_2} demonstrates that, even though intrinsically-deformed GCM states carry such a spurious coupling, it has no impact on the associated monopole strength function. In particular, while the full AMP accomplished via VAP-GCM eventually improve over GCM results the latter still deliver a meaningful approximation of the former\footnote{As discussed above in connection with Fig.~\ref{fig:Si28_0}, in $^{28}$Si the GCM monopole strength actually provides a quantitatively satisfactory approximation to the VAP-GCM one.}. Contrarily,  the subtraction of the spurious coupling strongly corrects the PAV-GCM strength function as seen in Fig.~\ref{fig:Si28_3} such that the sub-PAV-GCM monopole response becomes consistent with the VAP-GCM one. In fact, the sub-PAV-GCM monopole response remains very close to the original GCM one, i.e. once the spurious component is removed, the effect of the AMP is underestimated when performed {\it a posteriori}.


\section{Conclusions}
\label{sec:pav_gcm_summary}

The impact of angular momentum projection (AMP) on the monopole and quadrupole responses of doubly open-shell nuclei has been investigated within the frame of the (projected) generator coordinate method ((P)GCM) based on intrinsically deformed mean-field states. More specifically, the objective was to investigate whether the AMP can be safely performed {\it a posteriori}, i.e. solving the secular equation for intrinsic states within the GCM and projecting the solutions on good angular momentum only afterwards. To do so, results were confronted with results from full PGCM calculations where the AMP is performed {\it a priori}, i.e. where the secular equation is solved directly for good-symmetry states. 

Using $^{28}$Si as typical example, the angular momentum projection was shown to have a non-negligible impact on both the monopole and quadrupole responses in full PGCM calculations. First, the position of the dominant peaks are shifted up by about $1$\,MeV ($2$\,MeV) in the $J=0$ ($J=2$) channel. Second, while only the relative weight of certain monopole transitions are modified, quadrupole transitions are strongly suppressed. 

Next, the {\it a posteriori} angular-momentum restoration was shown to contaminate the monopole response with an unphysical coupling to the rotational motion, a result that is fully consistent with the one recently observed in (quasi-particle) random phase approximation calculations based on a deformed reference state~\cite{Porro23p}. Eventually, the present work based on the PGCM confirms that an {\it a priori} angular momentum restoration is necessary to handle consistently rotational and vibrational collective motions at the same time, which further pleads in favor of implementing the full-fledged projected (quasi-particle) random phase approximation in the future.

\section*{Acknowledgements}

Calculations were performed by using HPC resources from GENCI-TGCC (Contract No. A0130513012). A.P. was supported by the CEA NUMERICS program, which has received funding from the European Union's Horizon 2020 research and innovation program under the Marie Sk{\l}odowska-Curie grant agreement No 800945. A.P. and R.R. are supported by the Deutsche Forschungsgemeinschaft (DFG, German Research Foundation) – Projektnummer 279384907 – SFB 1245. R.R. acknowledges support though the BMBF Verbundprojekt 05P2021 (ErUM-FSP T07, Contract No. 05P21RDFNB). M.F. is supported by the CEA-SINET project.

\section*{Data Availability Statement}
This manuscript has no associated data or the data will not be deposited.

\bibliographystyle{apsrev4-1}
\bibliography{biblio_PGCM4}

\end{document}